\documentclass[final]{raa}            

\usepackage{graphicx,times}             
\usepackage{natbib}
\usepackage{amssymb,amsmath}
\bibpunct{(}{)}{;}{a}{}{,}

\newcommand{\cm}{{~\rm cm}}
\newcommand{\km}{{~\rm km}}
\newcommand{\s}{{~\rm s}}

\newcommand{\K}{{~\rm K}}
\newcommand{\erg}{{~\rm erg}}
\newcommand{\yr}{{~\rm yr}}

\newcommand{\pc}{{~\rm pc}}

\newcommand{\AU}{{~\rm AU}}
\newcommand{\mum}{{~\rm \mu m}}

\begin{document}

   \title{The circumstellar matter of type II intermediate luminosity optical transients (ILOTs)
}

   \volnopage{Vol.0 (20xx) No.0, 000--000}      
   \setcounter{page}{1}          

   \author{Noam Soker
      \inst{1,2}
   }

   \institute{Department of Physics, Technion, Haifa, 3200003, Israel;  soker@physics.technion.ac.il {\it soker@physics.technion.ac.il}\\
        \and
             Guangdong Technion Israel Institute of Technology, Shantou 515069, Guangdong Province, China\\
\vs\no
   {\small Received~~20xx month day; accepted~~20xx~~month day}}

\abstract{
I find that a $\simeq 0.1-1 M_\odot$ outflowing equatorial dusty disk (torus)  that the binary system progenitor of an intermediate luminosity optical transient (ILOT) ejects several years to several months before and during the outburst can reduce the total emission to an equatorial observer by two orders of magnitude and shifts the emission to wavelengths of mainly $\lambda \ga 10 \mum$. This is termed a type II ILOT (ILOT~II). To reach this conclusion I use calculations of type II active galactic nuclei and apply them to the equatorial ejecta (disk/torus) of ILOTs II. This reduction in emission can last for tens of years after outburst. Most of the radiation escapes along the polar directions. The attenuation of the emission for wavelengths of $\lambda< 5 \mum$ can be more than three orders of magnitudes, and the emission at $\lambda \la 2 \mum$ is negligible. Jets that the binary system launches during the outburst can collide with polar CSM and emit radiation above the equatorial plane and dust in the polar outflow can reflect emission from the central source. Therefore, during the event itself the equatorial observer might detect an ILOT. I strengthen the  previously suggested ILOT II scenario to the event N6946-BH1, where a red giant star disappeared in the visible.
}

 \authorrunning{N. Soker}            
   \titlerunning{The CSM of type II ILOTs}  

   \maketitle

%
%


\section{Introduction}
\label{sec:intro}
    
This study aims at eruptive transients, sometimes referred to as gap transients (e.g., \citealt{Kasliwal2011, Blagorodnovaetal2017, PastorelloFraser2019}) because they have peak luminosities between luminosities of classical novae and typical luminosities of supernovae  (e.g. \citealt{Mouldetal1990, Bondetal2003, Rau2007, Ofek2008, Masonetal2010, Tylendaetal2013, Kasliwal2013,  Kaminskietal2018, Pastorelloetal2018, BoianGroh2019, Caietal2019, Jencsonetal2019, Kashietal2019, PastorelloMasonetal2019, Andrewsetal2020, Howittetal2020, Jones2020, Kaminskietal2020Nova1670, Klenckietal2020, Kaminskietal2021Nova1670}). I refer to transients that are powered by gravitational energy due to stellar merger, that includes the onset of a common envelope evolution or mass transfer (e.g.,  \citealt{Tylendaetal2011, Nandezetal2014, Kaminskietal2015b, Soker2016GEEI, MacLeodetal2017, Gilkisetal2019, Segevetal2019, YalinewichMatzner2019, Schrderetal2020, MacLeodLoeb2020}) as intermediate luminosity optical transients (ILOTs;  \citealt{Berger2009, KashiSoker2016, MuthukrishnaetalM2019}) \footnote{There are other terms to ILOTs, like  intermediate-luminosity transients, and different classifications to sub-classes (e.g., \citealt{KashiSoker2016}), such as luminous red novae and others (e.g. \citealt{Jencsonetal2019, PastorelloFraser2019}).}.  
Both mass transfer and stellar merger lead to high accretion rates that power the transient, the high-accretion-powered ILOT model \citep{KashiSoker2016, SokerKashi2016TwoI}. 
In rare ILOTs the low mass companion might be a planet (e.g., \citealt{RetterMarom2003, Bearetal2011}). 
 
Specifically, this study considers ILOT where before, during and/or after the outburst the system suffers a very high equatorial mass loss rate that obscures most of the ILOT radiation, mainly in the visible and at shorter wavelengths, from an equatorial observer. \cite{KashiSoker2017} presented this class of ILOTs and, following the nomenclature of active galactic nuclei (AGNs), termed this group \textit{Type II ILOTs} (ILOTs II). 
 Basically, because ILOTs are not terminal explosions,  on a long time-scale (years to tens of years) the luminosity of regular ILOTs (non-type II ILOTs) returns to its more or less pre-outburst value. ILOTs II are those ILOTs where the luminosity as an equatorial observer measures stays much below the pre-outburst luminosity for years to tens of years. This decline in luminosity is likely to start years to months before the outburst. As well, the dusty equatorial outflow makes the emission to be mainly in the IR.  

I follow many studies according to which ILOTs are powered by binary interaction (e.g., \citealt{Kashietal2010, Tylendaetal2011, McleySoker2014, Nandezetal2014, Kaminskietal2015a, Kaminskietal2015b,  IvanovaNandez2016, Pejchaetal2016a, Soker2016GEEI, Zhuetal2016, Blagorodnovaetal2017, MacLeodetal2017, MacLeodetal2018, Michaelisetal2018, PastorelloMasonetal2019}). 
 The type of binary interaction can be one of the following, each of which releases gravitational energy. (a) A merger process where one star is destroyed by the other star (e.g., \citealt{SokerTylenda2003} for V838~Mon). The surviving star accretes part of the mass of the destroyed star, where the rest is ejected. Ejection of mass is mainly in an equatorial outflow and/or in two jets perpendicular to the equatorial plane.  This process cannot repeat itself. (b) Mass transfer from a giant star to a more compact companion. The more compact companions accretes mass and might launch two opposite jets (e.g., \citealt{KashiSoker2010} for Eta~Carinae). The system might lose mass through the second Lagrangian point to form an equatorial outflow. Both stars might survive and the ILOT might repeat itself. (c) The system enters a common envelope evolution (e.g., \citealt{Howittetal2020}). This is likely to form a dense equatorial outflow. The common envelope evolution can repeat itself only if the companion exits the common envelope (\citealt{Gilkisetal2019} for a neutron star companion). Note that process (c) might come after process (b). 

I further assume, as did some earlier studies of ILOTs (e.g., \citealt{Pejchaetal2016a, Pejchaetal2016b, HubovaPejcha2019}) that the binary interaction ejects mass in the equatorial plane. However, I do not attribute most of the ILOT emission to mass collision in the equatorial plane (e.g., \citealt{Pejchaetal2016a, Pejchaetal2016b, MetzgerPejcha2017}), but rather to jets (e.g., \citealt{Kashi2010, SokerKaplan2020, Soker2020}). In the present study the main role of the equatorial gas is to absorb and re-emit the ILOT radiation.

The present method is to use results from studies of type II AGNs where many more observations exist, and to apply the results to the model of ILOTs II. 
In Section 2 I briefly mention why such an application of type II AGN models to ILOTs II is possible and in section 3 I discuss the result of this comparison. In section 4 I strengthen the claim of \cite{KashiSoker2017} that the transient event N6946-BH1 could be in principle an ILOT II, though the case is not closed yet. I summarise in section \ref{sec:summary}.

\section{Dust tori in AGNs and in ILOTs}
\label{sec:Thermal}
\subsection{Gas density}
\label{subsec:density}

Higher gas density supplies external pressure that increases the dust sublimation temperature and hence allows dust to survive closer to the central radiation source (e.g., figures 4 and 5 of \citealt{BaskinLaor2018}). \cite{BaskinLaor2018} take the typical number density of the broad line region of AGN to be $n \approx 10^{11} \cm^{-3}$.  
Interestingly, the density in the equatorial  pre-outburst outflow of ILOTs II in the relevant radii is about similar. The relevant typical radius $R$ should be larger than the minimum radius where dust survives $R_{\rm dust}$. As we show below in ILOTs this radius is $R_{\rm dust} \approx 10^{14} \cm \approx 10 \AU$. So I scale quantities with the radius $R =100 \AU$. 
 
I take the typical values from \cite{KashiSoker2017}, who aimed at explaining 6946-BH1 as ILOT II, for a strongly interacting binary system that ejects a pre-outburst outflow  of mass $M_{\rm ej,e}$ into a narrow disk-shaped outflow around the equatorial plane, i.e., within a half opening angle of $\alpha_{\rm e} \simeq 15^\circ-30^\circ$ measured from the equatorial plane. This pre-outburst  outflow corresponds to a solid angle of $\Omega_e = 4 \pi \delta_e$ with $\delta_e = \sin \alpha_{\rm e} \simeq 0.26-0.5$. For about a solar mass that the binary system ejects into a thin expanding disk (torus) at a velocity of $v_e$ during a time $\Delta t_e$ before the ILOT event, the electron density at a distance $r$ from the center  is about  
\begin{eqnarray}
\begin{aligned} 
n_e \simeq  & 1.5 \times 10^{11}   
\left( \frac{\alpha_{\rm e}}{15^{\circ}} \right)^{-1} \left( \frac {M_{\rm ej,e}}{1 M_\odot} \right)
\left(\frac{\Delta t_{\rm e}}{3 \yr}\right)^{-1}
\\ & \times
\left( \frac {v_{\rm e}}{100 \km \s^{-1}} \right)^{-1}
\left( \frac {r}{100 \AU } \right)^{-2} \cm^{-3}.
\label{eq:NeEquator}
\end{aligned}
\end{eqnarray}
For a disk expanding from the binary system with a radial velocity of $v_{\rm e} = 100 \km \s^{-1}$ and  with a perpendicular velocity of about the sound speed $C_{\rm s}\approx 15 \km \s^{-1}$, the half opening angle of the disk is $\simeq C_{\rm S}/v_{\rm e} \simeq 10^\circ$. This serves as the justification for the scaling with $\alpha_{\rm e} = 15^\circ$. 
 
\subsection{Inner radius of dusty disk (torus)}
\label{subsec:torus}

Dust grains efficiently absorb UV radiation. In AGN that emit mainly in the UV and visible \cite{Barvainis1987} derived for the inner boundary of the dusty torus a value of $R_{\rm A,in} \simeq 1.3 L^{1/2}_{A46} (T_{\rm d,s}/1500 \K)^{-2.8} \pc$,
where $L_{A46}$ is the AGN luminosity in units of $10^{46} \erg$ and $T_{d,s}$ is the dust sublimation temperature. 
According to \cite{LaorDraine1993}, however, the value of $R_{\rm A,in}$ can be a factor of several smaller. For spherical dust grains with radii of $a=0.25 \mum$ as in \cite{KashiSoker2017}, I find from \cite{LaorDraine1993} a radius $R_{\rm A,in}$ that is about an order of magnitude smaller. But this depends on grain composition, size, and the density (section \ref{subsec:density}).  
In applying this formula to ILOTS we need to consider the following (e.g., see \citealt{Tylenda2005} for the evolution of V838~Mon). ($i$) The typical emission of ILOTs is in the visible and IR, with less emission in the UV (e.g., for the transient N6946-BH1; \citealt{Adamsetal2017}). ($ii$) The peak luminosity lasts for months and less, and therefore if there is dense dust in the CSM it might be that not all of it has time to sublimate. This is because at temperatures just above $1500 \K$ sublimation time is weeks for small grains to over a year for large grains and high surrounding gas densities (e.g., \citealt{BaskinLaor2018}). For that, the luminosity in that formula should be some average luminosity, and not that at the peak. ($iii$) Giant stars with low effective temperatures known to have dust close to them. 

The relation that \cite{KashiSoker2017} derive for ILOTs II (their equation 11) is close to the results of \cite{LaorDraine1993} for AGNs. The radius above which dust survives for a long time is 
\begin{equation}
R_{\rm dust}  \approx  20 
\left(\frac {L_{\rm I}} {3 \times 10^5 L_\odot} \right)^{1/2} 
\left(\frac {T_{\rm d,s}}{1500 \K} \right)^{-2.5} \AU ,  
\label{eq:Rdin}
\end{equation}
where ${L_{\rm I}}$ is the ILOT luminosity that now includes all the spectrum. 
The actual inner radius at which dust survives is smaller than this value because the high gas density (equation \ref{eq:NeEquator}) increases the sublimation temperature and the life of the dust grains \citep{BaskinLaor2018}. 
I take it that dust can exist in large quantities from $R_{\rm dust} \approx 5-10 \AU \approx 10^{14} \cm$ and outward. 

For an ILOT to be highly obscured by a dense equatorial dust the high mass loss rate  of the pre-outburst outflow  should start earlier than a time period of $\Delta t_{\rm e} \ga t_{\rm dust}$ before the outburts, where $t_{\rm dust}$ is the time for the equatorial  pre-outburst outflow  to reach the radius $R_{\rm dust}$. So the condition on the beginning of the pre-outburst high mass loss rate is  
\begin{equation}
\Delta t_{\rm e} \ga t_{\rm dust} \approx 0.5  
\left( \frac{R_{\rm dust}}{10 \AU} \right) 
\left( \frac {v_{\rm e}}{100 \km \s^{-1}} \right)^{-1} \yr.
\label{eq:Deltate}
\end{equation} 
Namely, massive equatorial  pre-outburst outflow   that starts about several months and longer before the main ILOT outburst can form a dust torus (disk) that might obscure the ILOT and/or the central remnants from an equatorial observer, i.e., turn the ILOT to ILOT II. 

\subsection{Geometry and expansion}
\label{subsec:Geometry}

There are significant qualitative differences between AGNs and ILOTs. While in AGNs the material in the obscuring torus/disk origin is from large distances in the galaxy and the torus neither expands or contracts much, in ILOTs the material in the disk originates in the binary interaction and flows radially outward. Therefore, some properties of the expanding disk change with time. 
   
\begin{figure}[t]
	\centering
	\hspace{1cm}
\includegraphics[trim=23cm 19cm 20cm 2cm ,clip, scale=0.52]{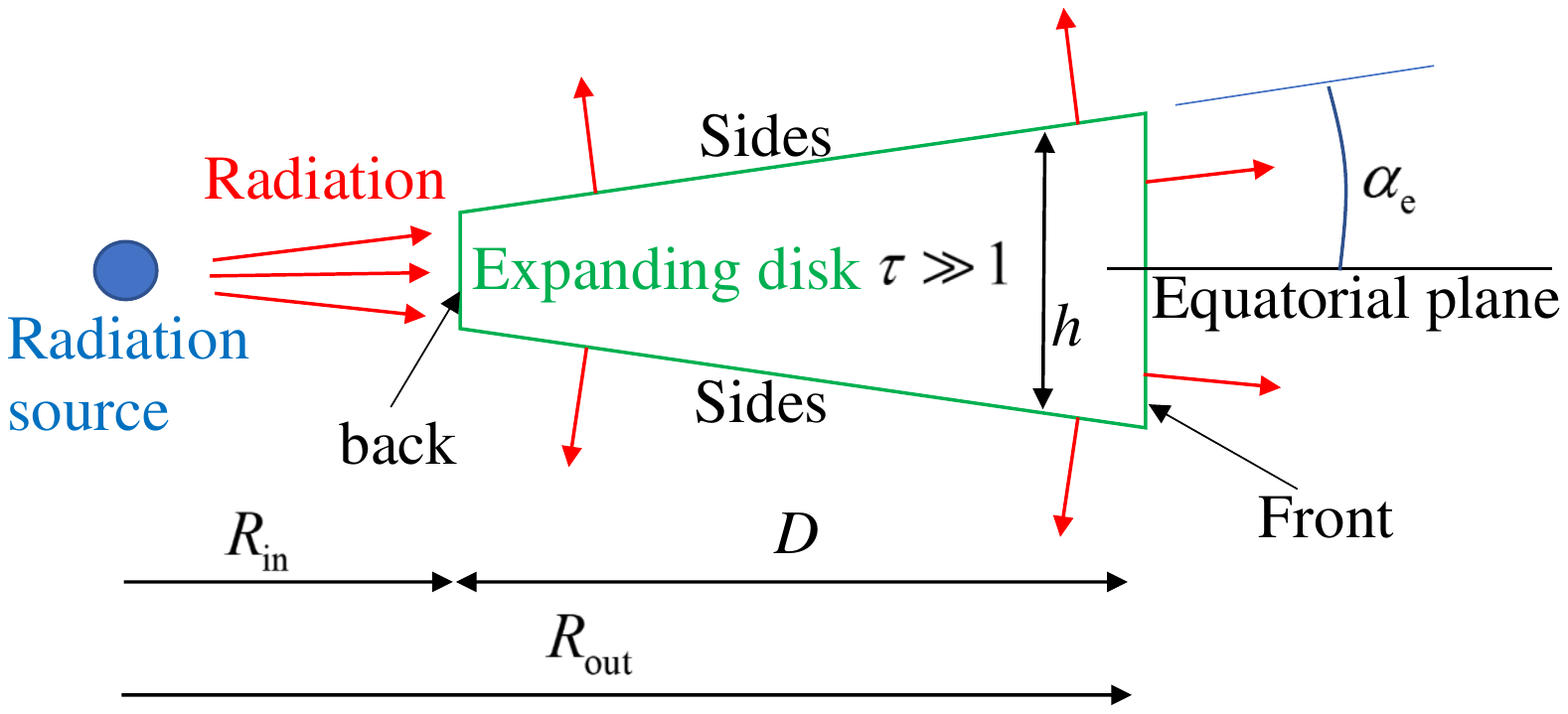}
\caption{A Schematic drawing of the expanding disk (torus) in the meridional plane and showing only one half of the meridional plane. Red arrows depict radiation. The $r$ coordinate is in the radial direction from the radiation source at the center, and the $z$ coordinate is perpendicular to the orbital plane. }
	\label{fig:schecmatic}
\end{figure}

I consider the ejection of an equatorial gas with a half opening angle of $\alpha_{\rm e}$ during the time period from $\Delta t_{\rm e}$ before outburst to the end of the outburst.  This gas includes the pre-outburst outflow and possibly additional outburst ejecta that the system ejects during the outburst.  I schematically present the expanding disk (torus) in Fig. \ref{fig:schecmatic}. 
 
In principle there are two limiting cases. In one there is a homologous expansion, namely, the radial velocity is proportional to the distance from the center. This might be the case in an `explosive' mass ejection in the equatorial plane. The other case is of a mass loss at a constant terminal velocity. Here I require the dense equatorial  pre-outburst outflow  to start months to years before the outburst, so I present a quantitative derivation for a mass loss at a constant terminal velocity $v_{\rm e}$. If the outburst itself ejects additional equatorial mass,  the outburst ejecta,  then the extinction toward an equatorial observer is larger even. The expansion of  the outburst ejecta,  might be close to homologous. 

Consider then a time $t$ post-outburst. The outer boundary of the disk, the `front', is at 
\begin{equation}
R_{\rm out} = R_{\rm in} +D = 211
\left( \frac{t + \Delta t_{\rm e}}{10 \yr}  \right)
\left( \frac {v_{\rm e}}{100 \km \s^{-1}} \right) \AU,
\label{eq:Rout}
\end{equation}
where $D$ is the length of the disk in the radial direction and $R_{\rm in}= v_{\rm e} t$ is the inner boundary, the `back', of the expanding disk. 
The full width of the thin disk at its middle is 
$H \simeq (R_{\rm in}+0.5D) \tan \alpha_{\rm e}$. 

For usage in section \ref{sec:equatorial} I calculate the typical optical depths along the disk radial direction and perpendicular to the orbital plane. The optical depths depend on the density $\rho(r,z)$, where $z$ is the coordinate along the direction perpendicular to the orbital plane, and on the opacity $\kappa$ inside the disk. The density in the disk depends on the mass loss rate  of the pre-outburst outflow  as function of time. For the crude estimates of the present study I take the  pre-outburst outflow  mass loss rate to be constant with time, and so $\rho(r) = \rho(R_{\rm in})(r/R_{\rm in})^{-2}$. The opacity depends on density, temperature, dust content (composition) and the wavelength. I assume that within the disk the opacity depends only on the wavelength. The optical depth from radius $r$ inside the disk to its outer boundary is 
\begin{eqnarray}
\begin{aligned}
\tau_r(\lambda) = & 
\int^{R_{\rm out}}_{r} \kappa (\lambda) \rho dr^{\prime} 
\simeq \kappa (\lambda) \rho (R_{\rm in})
R_{\rm in} 
\\ & \times \left( 
\frac{R_{\rm in}}{r}- 
\frac{R_{\rm in}}{R_{\rm out}} \right) \equiv 
\tau_{\lambda,0} \left( 
\frac{R_{\rm in}}{r}- 
\frac{R_{\rm in}}{R_{\rm out}} \right) 
\label{eq:Taur1},
\end{aligned}
\end{eqnarray}
where the second equality defines $\tau_{\lambda,0}$.

For the optical depth perpendicular to the disk I assume a thin disk so that the boundaries (sides) are $\pm h/2 \simeq r \alpha_{\rm e}$. The optical depth is then  
\begin{eqnarray}
\begin{aligned}
\tau_z(\lambda) = &  \int^{-h/2}_{h/2} \kappa (\lambda) \rho dz \simeq 2 \alpha_{\rm e} \kappa (\lambda) \rho (R_{\rm in}) R_{\rm in} \\ & \times
\left(\frac{R_{\rm in}}{r} \right)  .
\label{eq:Tauz1}
\end{aligned}
\end{eqnarray}
 At early times  most of the contribution to the opacity in the radial direction comes from regions of $r \ll R_{\rm out}$, and so  
\begin{equation}
\frac {\tau_z(\lambda)} {\tau_r(\lambda)} \simeq 2 \alpha_{\rm e} \left( 
1- \frac{r}{R_{\rm out}} \right) ^{-1} \approx 0.5 \left(\frac {\alpha_{\rm e}}{15^\circ} \right); \qquad t \ll \Delta t_{\rm e} . 
\label{eq:TauRatio}
\end{equation}
For a post-outburst of several times the mass loss period before outburst, $t \simeq {\rm several } \times \Delta t_{\rm e}$,  I take the contribution from the inner region of the disk where $r\simeq R_{\rm in} \simeq v_{\rm e} t$, and so (scaling for $t=  5\Delta t_{\rm e}$) 
\begin{equation}
\frac {\tau_z(\lambda)} {\tau_r(\lambda)} 
\approx 3 \left(\frac {\alpha_{\rm e}}{15^\circ} \right)
\left( \frac{t+\Delta t{\rm e}}{6 \Delta t_{\rm e}}  \right) ; 
\qquad t \ga \Delta t_{\rm e}. 
\label{eq:TauRatiolate}
\end{equation}

 Equations (\ref{eq:TauRatio}) and (\ref{eq:TauRatiolate}) refer only to the pre-outburst outflow. As stated earlier, an outburst ejecta that bounds the pre-outburst outflow from inside adds more mass that is more likely to extend to lower velocities. This implies a smaller inner radius even, and a lower value in equation (\ref{eq:TauRatiolate}) .   
It is possible that the binary interaction leads to the ejection of a spiral-disk (e.g., \citealt{HubovaPejcha2019, Kimetal2019}). In such a case the perpendicular optical depth in the zone between the spiral arms is very low. Effectively, this reduces even more the value of $\tau_z(\lambda)$. 
 Overall, I crudely take $\tau_z(\lambda)/ {\tau_r(\lambda)} \approx 1$ when I scale quantities, but I note that for thicker disks and at late times this ratio is larger, and might be as large as $\approx 10$ for, e.g., $\alpha_{\rm e} \simeq 30^\circ$ and $t \simeq 10 \Delta t_{\rm e}$. 

I scale the relevant quantities of equations (\ref{eq:Taur1}) and (\ref{eq:Tauz1}) for Thomson cross section of electron scattering (as, e.g., \citealt{PierKrolik1992} do)  
\begin{eqnarray}
\begin{aligned}
\tau_{\rm T,0} \equiv  & \kappa_{\rm T} \rho (R_{\rm in} ) R_{\rm in} \simeq 
 150
\left( \frac{\alpha_{\rm e}}{15^{\circ}} \right)^{-1}
\left( \frac {M_{\rm ej,e}}{1 M_\odot} \right)
\\ & \times
\left(\frac{\Delta t_{\rm e}}{3 \yr}\right)^{-1}
\left( \frac {v_{\rm e}}{100 \km \s^{-1}} \right)^{-1}
\left( \frac {R_{\rm in}}{100 \AU } \right)^{-1} .
\label{eq:TauT0}
\end{aligned}
\end{eqnarray}
where I used equation (\ref{eq:NeEquator}). 
With a radial velocity of $v_{\rm e}=100 \km \s^{-1}$ the time after ILOT outburst when the inner boundary of the disk is at $R_{\rm min} = 100 \AU$ is $t= 4.7 \yr$. 

To calculate the Thomson optical depth along the radial direction we need to multiply by the parenthesis in equation (\ref{eq:Taur1}). For example, consider a time $t=30 \yr$ after an ILOT outburst that suffered 3 years of pre-outburst high mass loss rate. For $v_{\rm e} = 100 \km \s^{-1}$ the disk radial boundaries are at $R_{\rm in} \simeq 630 \AU$ and at $R_{\rm out} = 1.1 R_{\rm in}$. Keeping other parameters as in the equations above, the radial Thomson optical depth of the entire disk, i.e., $r=R_{\rm in}$, is $\tau_{{\rm T},R_{\rm in}} \simeq 2$. 
We can combine equations (\ref{eq:Rdin}), (\ref{eq:Taur1}) and (\ref{eq:TauT0}) to scale the Thomson radial optical depth   
\begin{eqnarray}
\begin{aligned}
\tau_{{\rm T},r}  \simeq    &  2
\left( \frac{\alpha_{\rm e}}{15^{\circ}} \right)^{-1}
\left( \frac {M_{\rm ej,e}}{1 M_\odot} \right)
\left( \frac {v_{\rm e}}{100 \km \s^{-1}} \right)^{-2}
\\ & \times
\left(\frac{t+ \Delta t_{\rm e}}{33 \yr}\right)^{-1}
\left(\frac{t}{30 \yr}\right)^{-1}
 .
\label{eq:TauTr}
\end{aligned}
\end{eqnarray}

At this late time and for the parameters I use here, the Thomson optical depth in the disk perpendicular to the orbital plane is $\tau_{{\rm T},z} \simeq 5 \tau_{{\rm T},r}$ (by equation \ref{eq:TauRatiolate}). However, if the mass ejection at the outburst itself is substantial, the outburst ejecta, I expect a more homologous outflow in the inner part of the disk. Namely, the disk extends to much closer distance to the center. I will therefore continue to assume $\tau_{\rm T,z} \simeq  \tau_{\rm T,r}$ at all times.  

Finally, I note that because there is dust in the disk, the optical depth at the visible and IR bands are much larger. For the same parameters as in equation (\ref{eq:TauTr}) I find the radial optical depths at $\lambda = 1.25 \mum$ and $\lambda = 9.7 \mum$, that some other studies use,  to  be $\tau_{r}(1.25, \mum) \simeq 400$ and $\tau_{r} (9.7 \mum) \simeq 80$, respectively. For the visible band these parameters give $\tau_{V,r} \simeq 1500$. So after tens of years the disk will be optically thick even for an ejected mass that is an order of magnitude lower than the value I use for scaling (i.e., for only $M_{\rm ej,e} = 0.1 M_\odot$; see also \citealt{KashiSoker2017}). 
     
\section{Extinction toward an equatorial observer}
\label{sec:equatorial}
 
Because of the the high complexity of ILOTs due to rapidly varying luminosity, and because there are no observations of ILOTs II to guide the calculations (but see section \ref{sec:N6946-BH1} below), in this study I do not perform radiative transfer calculations. I rather use results from calculations of AGNs where there are plenty of observations of Type II AGNs (e.g., \citealt{Fritzetal2006}). In that I set the stage for similar calculations of ILOTs II in future studies. 

I start with the paper of \cite{PierKrolik1992} who calculated the emission to observers at different inclination angles with respect to a geometrically thick torus around the central radiation source. Their torus has a constant width $h$, i.e., a rectangle cross section in the meridional plane. To the present study only the observer in the equatorial plane that \cite{PierKrolik1992} considered is relevant. However, I note that because I take the disk to have a conical cross section in the meridional plane, in the present case all observers within an angle $\alpha_{\rm e}$ from the equatorial plane observe about the same emission. \cite{PierKrolik1992} considered tori that cover a large solid angle with their inner boundary. Namely, they considered only cases with $R_{\rm in} \le h$. 

For an equatorial observer they found that the peak luminosity decreases with increasing ratio of $R_{\rm in}/h$, where in their cases  $R_{\rm in} \le h$. For example, for Thomson optical depths of $\tau_{\rm T,r}=\tau_{\rm T,z}=1$, they find the peak values of $\lambda F_\lambda$, where $F_{\lambda}$ is the luminosity per unit solid angle per unit wavelength, to be $\lambda F_\lambda (0.1) \simeq 0.2 L/4 \pi$, 
$\lambda F_\lambda (0.3) \simeq 0.06 L/4 \pi$,  and
$\lambda F_\lambda (1) \simeq 0.02 L/4 \pi$, for $R_{\rm in} = 0.1 h$, $R_{\rm in} = 0.3 h$, and $R_{\rm in} =  h$, respectively, where $L$ is the luminosity of the central source. 
As said, in the present study where I use a disk (torus) with a conical cross section the most relevant case from these is the case of $R_{\rm in} =  h$.  The reason is that \cite{PierKrolik1992} considered only tori with $R_{\rm in} \le h$, while I consider disks with $  R_{\rm in} \ga h $.  
In these three cases the curve of $\lambda f_\lambda$ has double-peaks at  $\lambda \simeq 7 \mum$ and $\lambda \simeq 15 \mum$. 
 
A key feature of the results of \cite{PierKrolik1992} is that the disk diverts radiation from the equatorial direction to directions with larger angles to the equatorial plane. For that, one must not use a slab approximation. The usage of a slab to calculate the extinction of the disk to equatorial observers of ILOTs II will give wrong results.
This diversion of equatorial to polar emission increases with decreasing ratio of perpendicular to radial optical depth $\tau_z/\tau_r$ (also, e.g.,  \citealt{Stalevskietal2012}). 
Another feature of these studies of type II AGNS is that the disk shifts the radiation to longer wavelengths that increase on average with increasing optical depth in the radial direction $\tau_r$.
 
Later studies reach similar conclusions, namely that a torus (disk) around an AGN can substantially reduce the flux for $\lambda \la 10 \mum$, for tori with smooth dust distribution (e.g., \citealt{GranatoDanese1994}) or for tori made of dense clouds (e.g., \citealt{Honigetal2006, Nenkovaetal2008}). 
 
\cite{Nenkovaetal2008} for example, studied disks with a conical cross section. Their results substantially strengthen the key feature of \cite{PierKrolik1992}. In particular, they show that narrow disks with $\alpha_e \la 30^\circ$ divert a large fraction of the equatorial emission to larger angles (their figure 8).  For their model of a disk with a radial optical depth in the visual of $\tau_{V,r}=300$ (compare to equation \ref{eq:TauTr} that has $\tau_{V,r} \simeq 1500$) they find the peak value of $\lambda F_{\lambda}/(L/4 \pi)$ to be $0.4$, $0.2$, $0.08$, and $0.015$ for disks with a half opening angle of $\alpha_e=60^\circ$, $45^\circ$, $30^\circ$, and $15^\circ$, respectively. Only about 10\% of the total radiated energy to the equatorial direction is at wavelength of $\lambda < 5 \mum$. The ratio between the emission to the polar directions (high angles close to the polar directions) to the emission to the equatorial directions increases as the disk narrows. This clearly demonstrates how thin disks, $\alpha_{\rm e} \la 30^\circ$, efficiently divert equatorial emission to polar emission.  
From the results of \cite{PierKrolik1992} and 
\cite{Nenkovaetal2008} I find that very crudely, keeping scaling of variables to power of 10 and exponents to multiplies of $1/2$, I can write for the peak of $\lambda F_{\lambda}$ in the equatorial plane 
\begin{eqnarray}
\begin{aligned}
& \eta_{\rm p} \equiv \left( \frac{\lambda F_{\lambda}}{L/4 \pi} \right)_{\rm peak}
 \approx 0.01
\left( \frac{\sin \alpha_e}{\sin 15^\circ} \right)^2
\left( \frac{\tau_{{\rm T},r}}{1} \right)^{-1/2}
\\ &  \times
\left( \frac{\tau_{{\rm T},z}}{\tau_{{\rm T},r}} \right)^{3/2}
\approx 0.01
\left( \frac{\sin \alpha_e}{\sin 15^\circ} \right)^2
\left( \frac{\tau_{V,r}}{10^3} \right)^{-1/2}
\left( \frac{\tau_{V,z}}{\tau_{V,r}} \right)^{3/2},
\label{eq:LambdaF}
\end{aligned}
\end{eqnarray}
where the first line is scaled with the Thomson optical depth in the radial direction and the second line with the radial optical depth in the visual. This very crude expression captures also the equatorial emission that \cite{Stalevskietal2012} present in their figure 4.  

Although being a very crude approximation to the numerical results cited above, equation (\ref{eq:LambdaF}) nonetheless captures the key properties of an optically thick disk in the radial direction, $\tau_{V,r} \ga 100$. In particular, an optical depth in the perpendicular direction that is lower than the radial optical depth, $\tau_z < \tau_r$,  allows a very efficient diversion of equatorial radiation to radiation into other angles. 
 Note though that in cases of not-very-thin disks and at late times $\tau_z > \tau_r$ (equation \ref{eq:TauRatiolate}). 

The reduction in the flux is accompanied by the shifting of the peak of $\lambda F_\lambda$ to longer wavelength. For the relevant optical depth to this study, i.e., $\tau_{{\rm T}, r} > 0.1$ (corresponding to a radial opacity in the visible of  $\tau_{V, r} > 100$) I find that the total flux at wavelength of $\lambda < 5 \mum$ is less than 10\% of the flux. 
Very crudely I express this as  
\begin{eqnarray}
\begin{aligned}
\frac{L(\lambda< 5 \mum)}{L} & \la 10^{-3} 
\left( \frac{\sin \alpha_e}{\sin 15^\circ} \right)^2
\left( \frac{\tau_{{\rm T},r}}{1} \right)^{-1/2}
\\ & \times
\left( \frac{\tau_{{\rm T},z}}{\tau_{{\rm T},r}} \right)^{3/2}.
\label{eq:L5}
\end{aligned}
\end{eqnarray}

I now use equation (\ref{eq:TauTr}) to substitute for the radial Thomson optical depth in equation (\ref{eq:LambdaF}) and take $\lambda F_\lambda$ at peak to crudely be the total luminosity per unit solid angle. Rounding again to power of 10 the different values, I derive the following very crude relation between the luminosity as inferred by an equatorial observer, $L_{\rm e}$, and the total ILOT luminosity. 
\begin{eqnarray}
\begin{aligned}
& \frac{L_{\rm e}} {L_{\rm I}}
 \approx 0.01
\left( \frac{\sin \alpha_e}{\sin 15^\circ} \right)^{5/2}
\left( \frac{\tau_{{\rm T},z}}{\tau_{{\rm T},r}} \right)^{3/2}
\left( \frac {M_{\rm ej,e}}{1 M_\odot} \right)^{-1/2}
\\ & \times
\left( \frac {v_{\rm e}}{100 \km \s^{-1}} \right)
\left(\frac{t+ \Delta t_{\rm e}}{33 \yr}\right)^{1/2}
\left(\frac{t}{30 \yr}\right)^{1/2}. 
\label{eq:Ltot}
\end{aligned}
\end{eqnarray}
As before, the total luminosity in the wave band $\lambda<5\mum$ is less than 10\% of this value. 

I note that the disk obscures the central source only. A geometrically thin disk does not intercept radiation from an emission source above the plane along the polar directions.
Such a source can result from a dusty polar outflow that reflects light from the center (e.g., \citealt{KashiSoker2017}) or from jets that collide with a CSM on both sides of the plane (e.g., \citealt{Soker2020, SokerKaplan2020}).  The reflected light from the polar regions will be highly polarised. Namely, polarisation can reveal reflected light in ILOTs II.

\section{The possible Type II ILOT N6946-BH1}
\label{sec:N6946-BH1}
  
The source N6946-BH1 in the galaxy NGC~6946 experienced an outburst in 2009 March \citep{Gerkeetal2015}. There is a dispute on whether this transient event was a failed supernova \citep{Adamsetal2017, Basingeretal2020} or whether it was an ILOT II (\citealt{KashiSoker2017}, and more on the dispute there). 
    
According to \cite{Adamsetal2017} the pre-outburst star of N6946-BH1 was a red supergiant with a mass of $\approx 25 M_\odot$ and a radius of $\approx 2 \AU$. 
About three years before the outburst the visible luminosity started to decrease. On the other hand, the near IR luminosity (observed in $3.6\mum$ and $4.5 \mum$) increased slowly. In the ILOT II scenario this time marks the beginning of the high mass loss rate  pre-outburst outflow , namely, $\Delta t_{\rm e} \simeq 3 \yr$. In the one year-long outburst in the visible the star reached a peak luminosity of $> 10^6 L_\odot$. In a time scale of several months the visible luminosity dropped by a factor of several hundred. The star disappeared in the visible. 

\cite{KashiSoker2017} argue that an equatorial torus diverts most of the emission of the central source toward the polar directions, and introduced the term type II ILOT. They further attribute the extra outburst emission to dust in a polar outflow that reflects emission from the center. As the polar dust dispersed in the fast polar outflow and the luminosity of the central source returned to (almost) normal, the emission in the visible substantially decreased. 
The observed evolution of the near IR luminosity that \cite{Adamsetal2017} present in their figure 2 is qualitatively consistent with the ILOT II model that \cite{KashiSoker2017} propose (see figure 2 in \citealt{KashiSoker2017}) and that I study in more details in the present paper.

\cite{Adamsetal2017} calculate the energy that a slab transmits, and found it to be significant. Based on that, they argue against the ILOT II scenario for N6946-BH1. As I showed in earlier sections a slab geometry is likely to give wrong results. The reason is that one must use a disk geometry that allows a large fraction of the radiation to escape to the sides of the disk. 

Following the study of \cite{KashiSoker2017} and this study, I actually scaled earlier equations with parameters appropriate for N6946-BH1 (as there is no other examples of a possible ILOT II). Equation (\ref{eq:Ltot}) shows that for these parameters even in 2040 the total emission toward an equatorial observer will be a percent of the central source luminosity, most of it in wavelengths of $\lambda > 10 \mum$. The present upper measured emission from N6946-BH1 is $\approx 2000 L_\odot$ \citep{Adamsetal2017, Basingeretal2020}, but it does not include emission in wavelength longer than $5 \mum$.   

The luminosity in the band $\lambda < 5 \mum$ will be only $\approx 10^{-3}$ that of the central source, with practically no emission below $2 \mum$. At present, with $t=11 \yr$, even an equatorial mass of $0.1 M_\odot$ reduces emission to be much below the upper limit from observations. 

I conclude that the ILOT II scenario to explain the behavior of  N6946-BH1 is still viable.

\section{Summary}
\label{sec:summary}

I considered ILOTs that are driven by binary interaction. In many cases a strong binary interaction might eject a dense equatorial outflow (section \ref{sec:intro}). I concentrated on cases where the binary system ejects an equatorial mass of $M_{\rm ej,e} \ga 0.1 M_\odot$ within few years to few months before the main outburst. Dust can survive at large quantities outside a radius of $R_{\rm dust} \approx 10 \AU$ (equation \ref{eq:Rdin}). Outside this radius the density of the gas is similar to the typical density of the obscuring torus of type II AGNs (equation \ref{eq:NeEquator}). As well, the optical depth of the dusty outflow overlaps with optical depths of AGN tori (equation \ref{eq:TauT0}).     

The similar densities and optical depths of the disk that some ILOTs might eject to those of the disks/tori in type II AGNs brought me to use results from several studies of AGNs to the present study. From these previous results I derived a very crude expression for the peak value of $\lambda F_\lambda$ relative to the luminosity per unit solid angle of the central source for an equatorial observer ($\eta_{\rm p}$ in equation \ref{eq:LambdaF}).
 
I assumed that the ratio $\eta_{\rm p}$ crudely represents the luminosity as an equatorial observer would deduce relative to the ILOT luminosity of the central source, $L_{\rm e}/L_{\rm I}$. I used then the expression for the Thomson optical depth along the radial direction through the expanding disk that the progenitor of the ILOT ejected before the outburst (equation \ref{eq:TauTr}), and derived a very crude expression for the ratio $L_{\rm e}/L_{\rm I}$ (equation \ref{eq:Ltot}). This expression, though very crude, gives the correct value to an order of magnitude and captures the main process where the disk diverts equatorial radiation to larger angles. The geometry of the disk is crucial. For that, one cannot use a slab geometry to calculate how a dusty equatorial outflow obscures the ILOT from an equatorial observer. 

The main conclusion of this study is that a $\simeq 0.1-1 M_\odot$ outflowing equatorial dusty disk (torus) can reduce the total emission to an equatorial observer by two orders of magnitude. \cite{KashiSoker2017} termed this a type II ILOT. The radiation that the disk does emit in the equatorial direction is mainly in wavelengths of $\lambda \ga 10 \mum$. The radiation in the band of $\lambda < 5 \mum$ is less than $10\%$ of the total equatorial emission $L_{\rm e}$. 

The geometrically thin disk obscures only the central source and its vicinity.   
Jets that the binary system are likely to launch (e.g., \citealt{Soker2020, SokerKaplan2020}) during the outburst can reflect light from the central source  (\citealt{KashiSoker2017}) and/or collide with polar CSM and emit radiation far above the equatorial plane. As a result of that an equatorial observer might observe the ILOT outburst itself even in the visible. 
I used these result to strengthen the claim of \cite{KashiSoker2017} that the ILOT II scenario might account for the event N6946-BH1, where a red giant star disappeared in the visible.

\begin{acknowledgements}

I thank Ari Laor, Amit Kashi  and an anonymous referee  for very useful comments. 
This research was supported by a grant from the Israel Science Foundation (420/16 and 769/20) and a grant from the Asher Space Research Fund at the Technion.
\end{acknowledgements}

\label{lastpage}
\end{document}